\documentclass{article} 
\def\bilderscale{0.28}
\usepackage{epsf,graphicx,bildermacros_kleinert,amssymb,times,macros2e}
\newlength{\pcm}
\newlength{\pmm}
\makeindex

\newcommand{\half}{\frac{1}{2}}

\newcommand{\A}{\alpha}
\newcommand{\B}{\beta}

\newcommand{\E}{\varepsilon}

\newcommand{\R}{\mathbb{R}}

\newcommand{\rme}{{\mathrm{e}}}
\newcommand{\rmd}{{\mathrm{d}}} 

\newcommand{\be}{\begin{equation}}
\newcommand{\ee}{\end{equation}}
\newcommand{\bea}{\begin{eqnarray}}
\newcommand{\eea}{\end{eqnarray}}
\newcommand{\nn}{\nonumber}
\newcommand{\weiter}{\nonumber \\ & & }
\newcommand{\nicht}[1]{}

\newcommand {\eq}[1]{(\ref{#1})}
\newcommand {\Eq}[1]{Eq.\hspace{0.55ex}(\ref{#1})}
\newcommand {\Eqs}[1]{Eqs.\hspace{0.55ex}(\ref{#1})}
\newcommand {\Fig}[1]{figure~\ref{#1}}

\newcommand {\ds}{\displaystyle}
\newcommand {\tx}{\textstyle}

\newcommand {\ind}[1]{\mathrm{#1}}
\newcommand {\p}{\partial}

\newcommand{\bn}{:\hspace*{-0.5ex}}
\newcommand{\en}{\hspace*{-0.5ex}:}
\newcommand{\1}{\,{\bf 1}\,}

\newcommand{\setcurrentlabel}[1]{\def\@currentlabel{#1}}
\newbox{\expbox}
\newlength{\explength}
\newcommand{\EXPhelp}[6]{%
\sbox{\expbox}{\ensuremath{#4#1}}%
\settowidth{\explength}{\rotatebox{90}{\ensuremath{#4\left#5\usebox{\expbox}\right#6}}}%
\ensuremath{#4\left#5\usebox{\expbox}\right#6%
_{{#4\hspace*{0.4ex}}\hspace*{-0.22\explength}#2}%
^{{#4\hspace*{0.4ex}}\hspace*{-0.22\explength}#3}}}
\newcommand{\EXPdu}[3]{%
\mathchoice{\EXPhelp{#1}{#2}{#3}{\displaystyle}{<}{>}}%
{\EXPhelp{#1}{#2}{#3}{\textstyle}{<}{>}}%
{\EXPhelp{#1}{#2}{#3}{\scriptstyle}{<}{>}}%
{\EXPhelp{#1}{#2}{#3}{\scriptscriptstyle}{<}{>}}}
\newcommand{\EXP}[2]{%
\mathchoice{\EXPhelp{#1}{#2}{}{\displaystyle}{<}{>}}%
{\EXPhelp{#1}{#2}{}{\textstyle}{<}{>}}%
{\EXPhelp{#1}{#2}{}{\scriptstyle}{<}{>}}%
{\EXPhelp{#1}{#2}{}{\scriptscriptstyle}{<}{>}}}
\newcommand{\BRAhelp}[6]{%
\sbox{\expbox}{\ensuremath{#4#1}}%
\settowidth{\explength}{\rotatebox{90}{\ensuremath{#4\left#5\usebox{\expbox}\right#6}}}%
\ensuremath{#4\left#5\usebox{\expbox}\right#6%
_{{#4\hspace*{0.4ex}}\hspace*{-0.2\explength}#2}%
^{{#4\hspace*{0.4ex}}\hspace*{-0.2\explength}#3}}}
\newcommand{\BRA}[2]{%
\mathchoice{\BRAhelp{#1}{}{#2}{\displaystyle}{(}{)}}%
{\BRAhelp{#1}{}{#2}{\textstyle}{(}{)}}%
{\BRAhelp{#1}{}{#2}{\scriptstyle}{(}{)}}%
{\BRAhelp{#1}{}{#2}{\scriptscriptstyle}{(}{)}}}
\newbox{\atbox}
\newlength{\atlengtha}
\newlength{\atlengthb}
\newcommand{\AThelp}[4]{%
\sbox{\atbox}{\ensuremath{#3#1}}%
\settoheight{\atlengtha}{\ensuremath{#3\usebox{\atbox}}}%
\settodepth{\atlengthb}{\ensuremath{#3\usebox{\atbox}}}%
#3\addtolength{\atlengtha}{0.1ex}
#3\addtolength{\atlengthb}{0.75ex}%
\addtolength{\atlengtha}{\atlengthb}%
#1\rule[-\atlengthb]{0.1ex}{\atlengtha}_{\raisebox{0.12ex}%
{\ensuremath{\,#4#2}}}}%
\newcommand{\AT}[2]{%
\mathchoice{\AThelp{#1}{#2}{\displaystyle}{\scriptstyle}}%
{\AThelp{#1}{#2}{\textstyle}{\scriptstyle}}%
{\AThelp{#1}{#2}{\scriptstyle}{\scriptscriptstyle}}%
{\AThelp{#1}{#2}{\scriptscriptstyle}{\scriptscriptstyle}}}
\newlength{\picheight}%

\newbox{\picbox}
\newlength{\picwidth}

\renewcommand{\newblock}{}

\begin{document}

\title{\sffamily\bfseries\Large Principles of non-local field theories
and their application to polymerized membranes}
\index{polymerized membrane}\index{tethered membrane}
\index{self-avoiding membranes}
\index{non-local field-theories}
\author{\bf\large Kay J\"org Wiese\footnote{New adress: ITP, Kohn
Hall, UCSB, Santa Barbara, CA 93106-4030, USA; email: wiese@itp.ucsb.edu} \smallskip\\
\normalsize Fachbereich Physik, Universit\"at Essen,  45117 Essen,
Germany\\ \small E-mail: wiese@theo-phys.uni-essen.de}


\maketitle

\abstract{In these lecture notes, we give an overview about non-local field-theories
and their application to  polymerized
 membranes, i.e.\ membranes with a fixed internal connectivity. 
The main technical tool is the multi-local operator product
expansion (MOPE), generalizing ideas from local field theories
to the multi-local situation. \\
These notes are largely inspired by: 
Kay Wiese, ``Polymerized membranes, a review'', Domb and Lebowitz, eds.,\ %
Academic Press, London (2000). \\
Lectures held for the Graduiertenkolleg in Heidelberg, 
September 25-27.
\\
Dedicated to Hagen Kleinert at the occasion of his 60th birthday.\\
Published in Fluctuating Paths and Fields, W. Janke, A. Pelster,
H.-J. Schmidt and M. Bachmann (Eds.), World Scientific Singapore (2001)
}

\newsavebox{\GHxydbox}
\sbox{\GHxydbox}{\ensuremath{\ds\reducedbildheightrule{\GH}{0.7}_x\hspace{-0.2ex}\GH\reducedbildheightrule{\GH}{0.7}_y}}
\newcommand{\GHxyd}{\usebox{\GHxydbox}}
\newcommand{\GHxyt}%
{\reducedbildheightrule{\GH}{0.7}_x\hspace*{-0.23ex}\GH\reducedbildheightrule{\GH}{0.7}_y}

\section{Polymerized membranes}
\label{Polymerized membranes}
Mmembranes have attracted much interest during the last
years, especially due to their relevance for biological systems. 
Most of the membranes encountered there are fluid.
In this lecture I  shall concentrate on another fascinating
class of membranes, 
namely polymerized tethered also called solid membranes. 
These membrane  have a fixed and 
constant internal metric. They are realized in experiments
(e.g.\ the spectrin network of red blood cells
\cite{ElsgaeterStokkeMikkelsenBranton1986,%
SchmidtSvobodaLeiPetscheBermanSafinyaGrest1993}, or sheets
of graphite oxide \cite{HwaKokufutaTanaka1991,%
SpectorNaranjoChiruvoluZasadzinski1994}). 

A microscopic model  is given by the 
so-called ``spring and bead model''  
which consists  of balls (beads) which are connected by
springs and form a regular lattice. 
The difficulty is to incororate the self-avoidance of the beads,
i.e.\ the fact that beads cannot intersect.

\section{Representation as non-local field-theory}
\index{non-local field-theory}
\label{m:Definition of the model...}
\begin{figure} [t]
\epsfxsize=0.75\textwidth \centerline{\epsfbox{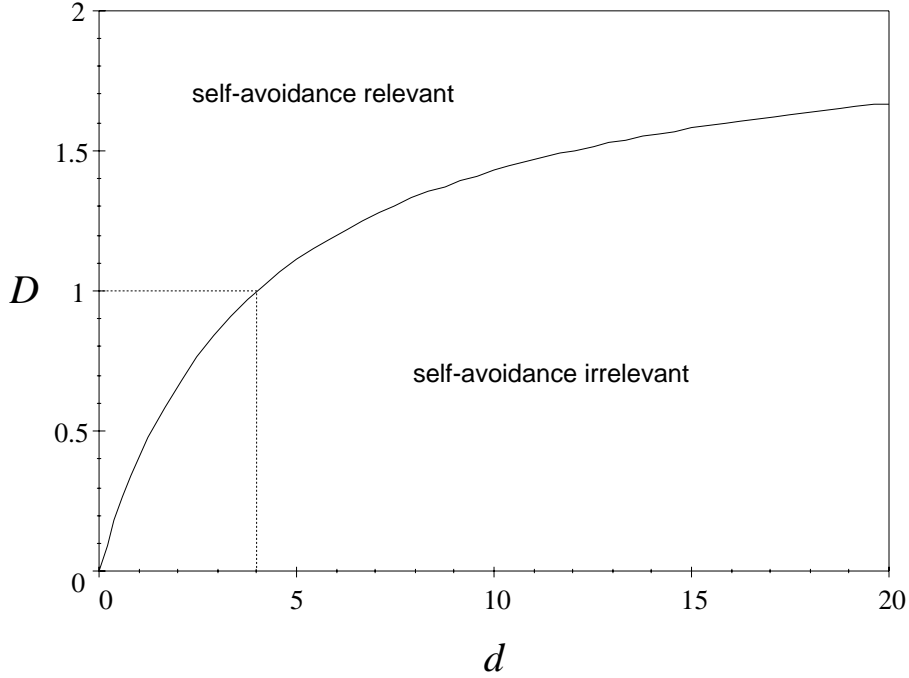}}
\caption{The critical curve $\E(D,d)=0$. The dashed line corresponds to the 
standard polymer perturbation theory, critical in $d=4$.}%
\label{m:kritdim}%
\end{figure}%

We start from the continuous model for a $D$-dimensional flexible polymerized
membrane introduced in \cite{AronowitzLubensky1987,KardarNelson1987},
and further studied in \cite{DDG1,DDG2,Duplantier1987,DuplantierHwaKardar1990}.
This model is a simple extension of the well known Edwards' model for
continuous chains. 
The membrane fluctuates in $d$-dimensional space.
Points in the membrane are labeled by coordinates $x\in\R^D$ and the 
configuration of the membrane in physical space is described by
the field $r:  x \in \R^D \longrightarrow r(x) \in \R^d$, i.e.\ from 
now on we note $r$ instead of $\vec r$. At high temperatures
the free energy for a configuration is given by the (properly 
scaled) Hamiltonian
\be
{\cal H}[r]
= \frac 1{2-D} \int_x \half \big(\nabla r(x)\big)^2 +
        b_0 \int_x \int_y \tilde\delta^d\big(r(x)-r(y)\big)
\ . 
\label{m:2ptact}
\ee
The integral $\int_x$ runs over $D$-dimensional space and $\nabla$ is the usual
gradient operator.
The normalizations are 
$
\int_ x:= {1 \over S_D} \int \rmd^Dx$ with  $S_D = 2 {\pi^{
D/2} \over \Gamma( D/2)} 
$
and
$\tilde \delta^d(r(x)-r(y)) = (4\pi)^{ d/2}\delta^d(r(x)-r(y)) 
$.
The latter term is normally used in  Fourier-representation
$
\tilde \delta^d(r(x)-r(y)) = 
 \int_p \rme^{ip[r(x)-r(y)]}$,
where the normalization of $\int_p$ is given by 
$ \int_p= \pi^{ -d/2} \int \rmd^dp $
to have
$\int_p \rme^{-p^2a} = a^{-d/2}$. 
All normalizations are chosen in order to simplify the calculations. 
%
The first term in the Hamiltonian 
is a Gaussian elastic  energy which is known to describe the
free ``phantom" surface.
The interaction term corresponds (for $b_0>0$) to a weak repulsive interaction
upon contact. 
The expectation values of physical observables are obtained by performing the
average over all field-configurations $r(x)$ with the Boltzmann
weight $e^{-{\cal H}[r]}$. This average can not be calculated 
exactly, but one can expand about the configurations of a
phantom, i.e.\ non-interacting surface. 

Such a perturbation theory is constructed by performing the series expansion in
powers of the coupling constant $b$.
This expansion suffers from ultraviolet  (UV) divergences which have to
be removed by renormalization and which are treated by dimensional 
regularization, i.e.\ analytical continuation in $D$ and $d$.
Long-range infrared (IR) divergences also appear.
They can be eliminated by using a finite membrane, or by studying translationally
invariant observables, whose perturbative expansion is also IR-finite in the
thermodynamic limit (infinite membrane).
Such observables are ``neutral'' products of vertex operators 
\be
{\cal O}=\prod_{a=1}^N {\rme}^{i k_a r(x_a)}\ , \qquad
\sum_{a=1}^N k_a\,=\,0
\ .
\label{m:prodvo}
\ee

Let us now analyze the theory by power-counting. We use internal 
units $\mu\sim 1/x$, and note  $\left[ x \right]_x =1$, 
and $\left[ \mu \right]_x =-\left[ \mu \right]_\mu =-1$ . The dimension 
of the field and of the coupling-constant are:
\be
 \nu :=\left[ r\right]_x  ={2-D\over 2}\ ,\quad \E:=\left[ b_0 \right]_\mu =2D-  \nu  d
\label{m:dimrg}
\ .
\ee
The interaction is relevant for $\E>0$, see figure \ref{m:kritdim}. 
Perturbation theory is then expected to  be UV-finite except for
subtractions associated to relevant operators. We shall come back to
this point later.

For clarity, we represent graphically the different interaction terms
which have to be considered.
The local operators are
\bea
1&=&\mbox{\bf 1}\ , \qquad 
\half\big(\nabla r(x)\big)^2=\GO 
\ .
\label{m:1ptops}
\eea
The bi-local operator, the dipole, is 
\bea
\tilde \delta^d\big( r(x)-r(y)\big)&=&\GB
\ .
\label{m:2ptops}
\eea
The expectation-value of an observable is
\be
	\EXP{{\cal O}[r]}{b} = \mathrm{Norm}
{\ds\int {\cal D}[r]\, {\cal O}[r]\,\mbox{\rm e}^{-{\cal H} \left[ r\right] }} \ .
\label{m:expectation}
\ee
with $\mathrm{Norm}$ chosen such that $\EXP{{\bf 1}}{b}=1$.
Perturbatively, all expectation-values are taken with respect
to the free theory (again normalized s.t. $\EXP{{\bf 1}}{0}=1$):
\be
\EXP{{\cal O}[r]}0=\mathrm{Norm}'
{\ds\int {\cal D}[r]\, {\cal O}[r]\,\mbox{\rm e}^{-\frac{1}{2-D} \int_x \half(\nabla r(x))^2}}
\label{m:free expectation}
\ .
\ee
A typical term in the expansion of (\ref{m:expectation}) is 
\be
\left(-b Z_b \mu^\E\right)^n \int\!\!\!\int \ldots \int\!\!\!\int \EXPdu{ {\cal O} \GB \ldots \GB }0{\ind{c}}
\ ,
\ee
where the integral runs over the positions of all   dipole-endpoints.

\section{Locality of Divergences and 
the multilocal operator product expansion (MOPE\index{MOPE})}
\index{locality of divergences}
\label{m:MOPE}
The next step to show ist that all
 divergences are short distance
divergences.
Note that even for massless theories and in the absence of 
IR-divergences, this is not trivial. 
Divergences could as well appear, when some of the distances
involved become equal, or multiple of each other. 
A simple counter-example is the integral of 
$\big| |a|-|b|\big|^{-\nu d}$, where $a$ and $b$ are two of 
the distances involved. 
Due to a lack of space, we can not give the proof here.
The interested reader should consult \cite{WieseHabil}.

As in local field-theoies, divergences 
can be analyzed via an operator product expansion. 
 Intuitively, in the context of 
multilocal theories --~by which we mean that the interaction
depends on more than one point~-- we also expect multilocal
operators  to appear in such an operator product expansion,
which therefore will be called ``multi-local operator product
expansion'' (MOPE\index{MOPE}) \cite{DDG3,DDG4}. 
An example for such an operator product expansion is
\be
\epsfxsize=0.9\textwidth \parbox{0.9\textwidth}{\epsfbox{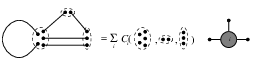}} 
\label{m:contraction} 
\ .
\ee
Let us explain the formula. We  consider $n$ dipoles
 (here $n=5$) and we separate the $2n$ end-points 
into $m$ subsets (here $m=3$) delimited by the dashed lines.
The MOPE describes how the product of these $n$ dipoles behaves when the points
inside each of the $m$ subsets are contracted towards a single point $z_j$.
The result is a sum over {\em multilocal} operators $\Phi_i(z_1,\ldots ,z_m)$,
depending on the $m$ points $z_1,\ldots ,z_m$, of the form
\be
\label{m:etheMope}
\sum_{i}C_i(x_1-z_1,\ldots)\,\Phi_i\left(  z_1, z_2, \ldots, z_m  \right)
\ ,
\ee
where the MOPE-coefficients $C_i(x_1-z_1,\ldots)$ 
depend only on the distances $x_l-z_j$ {\em inside each subset}.
This expansion is  valid as an operator-identity, i.e.\ inserted 
into any expectation value and in the limit of small distances between contracted points. No other operator should appear at the points
$z_1,\ldots, z_m$, towards which the operators are contracted. 
As the Hamiltonian (\ref{m:2ptact}) does not contain a mass-scale, the
MOPE-coefficients are homogeneous functions of the relative positions between the
contracted points, with the degree of homogeneity given by  dimensional analysis.
In the case considered here, where $n$ dipoles are contracted to an operator
$\Phi_i$, this degree is  $-n \nu d - [\Phi_i]_x$.
This means that
\be
C_i(\lambda(x_1-z_1) ,\ldots ) = \lambda^{-n \frac{2-D}2 d - [\Phi_i]_x} C_i(x_1-z_1 ,\ldots ) \ ,
\ee
where $[\Phi_i]_x$ is the canonical dimension of the operator $\Phi_i$ and 
$-d(2-D)/2$ is the canonical dimension of the dipole.

In order to evaluate the associated singularity, one finally has to integrate
over all relative distances inside each subset.
This gives an additional scale factor with degree $D(2n-m)$.
A singular configuration, such as in \Eq{m:contraction},
will be UV-divergent if this degree of divergence
\be
	D(2n-m)-n\frac{2-D}2d-\left[\Phi_i\right]_x \ ,
\ee
 is negative.
It is superficially divergent if the degree is zero and convergent otherwise.
The idea of renormalization
is {\em to remove exactly these superficially divergent contributions
recursively}.

\section{Evaluation of the MOPE-coefficients}
\label{m:Div and MOPE}
The MOPE gives a convenient and powerful tool to calculate the
dominant and all subdominant contributions from singular configurations.
In this section, we explain how to calculate the MOPE-coefficients
on some explicit examples. These examples will turn out to be 
the necessary diagrams at 1-loop order. 

In the following we shall use the notion of normal-ordering.
The first thing, which we use, is that 
\vspace*{-2mm}
\be \label{m:NO}
\bn \rme^{ikr(x)} \en \  = \rme^{ikr(x)}\ .
\ee 
Explicitly, tadpole-like contributions which are powers of  
$ \int \rmd^D p\, \frac{1}{p^2}$
are omitted. This is done via a finite part prescription
(analytic continuation, dimensional regularization),
valid for infinite membranes, for which the normal-order
prescription is defined. 
Let us stress that this is a pure technical trick, which 
can be cirumvented at the expense of more cumbersome
calculations.

The key-formula for all further manipulations is
\be \label{m:key}
\bn  \rme^{i k r(x)} \en \, \bn  \rme^{i p r(y)} \en\ 
=\ \rme^{kpC(x-y)} \bn  \rme^{i k r(x)}  \rme^{i p r(y)} \en
\ .
\ee
This can be proven as follows: Consider the (free) expectation
value of any observable $\cal O$ times the operators of \Eq{m:key}.
Then the left- and right-hand sides of the above equation read 
$
{\cal L} = \EXP{ {\cal O} \bn  \rme^{i k r(x)} \en \, \bn  \rme^{i p r(y)} \en\ }0 $ and 
${\cal R}= \rme^{kpC(x-y)}\EXP{ {\cal O} \bn  \rme^{i k r(x)}  \rme^{i p r(y)}\en }0$.
First of all, for ${\cal O}=1$, the desired equality of
 ${\cal L}={\cal R}$ holds, because
 $\EXP{\bn  \rme^{i k r(x)}  \rme^{i p r(y)}\en }0=1$
 and 
$\EXP{ \bn  \rme^{i k r(x)} \en \, \bn  \rme^{i p r(y)} \en}0=\rme^{kpC(x-y)}$.
Now consider a non-trivial observable $\cal O$, and contract 
all its fields $r$ with $\rme^{i k r(x)}$ or $\rme^{i p r(y)}$,
before contracting any of the fields $r(x)$ with $r(y)$. 
The result is a product of correlation-functions between the
points in $\cal O$ and $x$ or $y$, and these are equivalent
for both $\cal L$ and $\cal R$. However, contracting an 
arbitrary number of times $\rme^{ikr(x)}$, leaves the exponential
$\rme^{ikr(x)}$ invariant. Completing the contractions 
for ${\cal L}$ therefore
yields a factor of $\rme^{kpC(x-y)}$, and the latter one 
also appears in $\cal R$. Thus, 
the equality of $\cal L$ and $\cal R$ holds for all $\cal O$
and this proves
\Eq{m:key}.

Now proceed to the first explicit example, the contraction
of a single dipole with endpoints $x$ and $y$:
\be
 \GHxyd =\int_k \bn\rme^{i k r(x)}\en\, \bn\rme^{-ikr(y)}\en 
\ .
\ee
This configuration may have divergences when $x$ and $y$ 
come close together. Let us stress that  in contrast to 
$\phi^4$-theory, these divergences are not obtained as a finite sum
of products of correlators: Since $C(x-y)=|x-y|^{2-D}$, 
the latter is always well-behaved at $x=y$. The singularity 
only appears when summing an infinite series of diagrams as we
will do now. To this purpose, we first normal-order the two 
exponentials using \Eq{m:key}
\be
\int_k \,\bn\rme^{i k \left[r(x) -r(y)\right]}\en \rme^{-k^2 |x-y|^{2\nu}} \ .
\ee
Note that  the operators $\rme^{i k r(x)}$ and $\rme^{-ikr(y)}$
are free of divergences upon approaching each other, since 
no more contractions can be made. The divergence is 
captured in the factor $\rme^{-k^2 |x-y|^{2\nu}}$.
Therefore, we can expand the exponential $\bn\rme^{i k \left[r(x) -r(y)\right]}\en$ for small $x-y$ and consequently in 
 powers of $[r(x)-r(y)]$. This expansion is 
\be
\int_k \bn \left\{ \mbox{\bf 1} +i k \left[r(x) -r(y)\right] - \half\left(k 
\left[r(x) -r(y)\right] \right)^2\, + \ldots \right\}\en
 \rme^{-k^2 |x-y|^{2\nu}} 
\ .
\ee
We truncated the expansion after the third term. It will turn
out later that this is sufficient, since subsequent 
terms in the expansion are proportional to irrelevant operators
for which the integral over the  MOPE-coefficient
is UV-convergent.  

Due to  the symmetry of the integration over $k$ the term 
linear in $k$ vanishes. Also due to symmetry, 
the next term can be simplified with the result
\be
 \int_k \left[ \mbox{\bf 1} - \frac{k^2}{2d} \,:\left[r(x)-r(y)\right]^2:\, + \ldots \right]
 \rme^{-k^2 |x-y|^{2\nu}} \ .
\ee
Finally, the integration over $k$ can be performed. 
Recall that normalizations were chosen  such that
$\int_k \rme^{-sk^2}=s^{-d/2}$ to obtain
\be \label{m:symbmope1}
\GH = \MOPE{\GH}{\,\mbox{\bf 1}\,} \mbox{\bf 1}\, + 
\MOPE{\GH}{\GOalphabeta} \GOalphabeta \, + \ldots \ ,
\ee 
where we used the notation 
${\ds\GOalphabeta}=\half (\partial_\A r)(\partial_\B r)$
and the MOPE-coefficients (reminding Feynman's bra and ket notation)
\bea 
\MOPE{\GH}{\,\mbox{\bf 1}\,}&=& |x-y|^{-\nu d} \\
\MOPE{\GH}{\GOalphabeta} &=& - \frac 1 {2}
\,(x-y)_\A \,(x-y)_\B \,  |x-y|^{-\nu (d+2)} \ .
\label{m:symbmope2}
\eea
As long as the angular average is taken (and this will be the 
case when integrating the  MOPE-coefficient to obtain the divergence),
we can replace in  \Eq{m:symbmope1} $\ds\GOalphabeta$
by ${\ds\GO}:=\half (\nabla r)^2$ and \Eq{m:symbmope2}  by  
\be
\MOPE{\GH}{\GO} = - \frac 1 {2D}
 |x-y|^{D-\nu d} \ .
\ee

Next consider a real multi-local example of an
operator-product expansion, namely the contraction
of two dipoles towards a single dipole:
\be \label{m:(=|-)}
 {_{x+u/2} \atop ^{x-u/2}}  \scalebox{1.35}{\ensuremath{\ds\GM}}
  {_{y+v/2} \atop ^{y-v/2}} =
 \int_k  \rme^{i k \left[r(x+u/2) -r(y+v/2)\right]}
 \int_p\rme^{i p \left[r(x-u/2) -r(y-v/2)\right]}
\ .
\ee
This has to be analyzed for small $u$ and $v$, in order to 
control the divergences in the latter distances. As above, we 
normal-order operators which are approached, yielding
\bea
\rme^{i k r(x+u/2)} \rme^{i p r(x-u/2)} \, &=&\  
\bn \rme^{i k r(x+u/2)}\en \, \bn  \rme^{i p r(x-u/2)}\en \nn\\
&=& \ \bn \rme^{i k r(x+u/2)} \rme^{i p r(x-u/2)} \en 
\rme^{kpC(u)}
\ .
\eea
A similar formula holds when 
approaching $\rme^{-i k r(y+v/2)}$ and $ \rme^{-i p r(y-v/2)}$. 
\Eq{m:(=|-)} then becomes
\be 
\int_k\int_p  \bn\rme^{i k r(x+u/2)+ipr(x-u/2)}\en
 \ \bn\rme^{-ik r(y+v/2)-i p r(y-v/2)}\en\, \rme^{kp\left[ C(u)+C(v)\right]}
\ .
\label{m:6-}
\ee
In order to keep things as simple as possible, let us first 
 extract the leading contribution before analyzing subleading corrections. 
This leading contribution is obtained when expanding the exponential
operators (here exemplified for the second one) as
\be \label{m:6}
\bn \rme^{-i k r(y+v/2)} \rme^{-i p r(y-v/2)} \en \ =\  
\bn \rme^{-i (k+p) r(y)} \left( 1+ O(\nabla r) \right)\en 
\ee
and dropping terms of order $\nabla r$. This simplifies \Eq{m:6-} to
\be
\int_k\int_p  \bn\rme^{i (k+p)r(x)}\en\ \bn\rme^{-i (k+p)r(y)}\en\, \rme^{kp\left[ C(u)+C(v)\right]} \ .
\ee
In the next step, first $k$
and second $p$ are shifted 
$
k \longrightarrow k-p$  then $ p \longrightarrow p+\frac k2$.
The result is (dropping the normal-ordering according to \Eq{m:NO})
\be \label{m:some result}
\int_k  \rme^{ik\left[ r(x)-r(y)\right]}\, 
\int_p	\rme^{(\frac14k^2-p^2) \left[ C(u)+C(v)\right]} \ .
\ee
The factor of $\int_k  \rme^{ik\left[ r(x)-r(y)\right]}$ is 
again a $\delta$-distribution, and the leading term of the short distance
expansion of \Eq{m:some result}.  
Derivatives of the $\delta$-distribution appear when 
expanding $\rme^{(\frac14k^2-p^2) \left[ C(u)+C(v)\right]}$
in $k^2$; these are less relevant 
and only the first sub-leading term will be  displayed for illustration:
\bea
&&\int_k  \rme^{ik\left[ r(x)-r(y)\right]} 
\int_p	\rme^{-p^2 \left[ C(u)+C(v)\right]} \left( 1 +\frac{k^2}4 
\left[ C(u)+C(v)\right] + \ldots \right) \nn\\
&&\qquad =  \MOPE{\GM}{\GB} \GB + 
  \MOPE\GM\GC  \GC\ + \ldots \ ,
\label{m:(=|-) result}
\eea
where in analogy to \Eqs{m:symbmope1} to \eq{m:symbmope2}
\bea
  \MOPE\GM  \GB  &=&  \left[ C(u)+C(v)\right]^{-d/2}\ ,
 \nn\\
 \MOPE\GM  \GC  &=& \frac14 \left[ C(u)+C(v)\right]^{1-d/2}
\eea
with $\GB=\tilde \delta^d(r(x)-r(y))$ and  $\GC = (-\Delta_r) \tilde \delta^d(r(x)-r(y))$.

Let us already mention that the leading contribution proportional 
to the $\delta$-distribution will renormalize the coupling-constant,
and that the next-to-leading term is irrelevant and can be neglected. The same 
holds true for the additional term proportional to $(\nabla r)$
which was dropped in \Eq{m:6}.

There is one more possible divergent contribution at the 1-loop
level, namely $\FD$. We now show that the leading term of 
its expansion, which is expected to be  proportional to $\GB$, 
is trivial. To this aim consider
\bea \label{m:uiui}
&& \!\!\!\!\!\!\FD
\raisebox{-4.5\pmm}[0mm][0mm]{\makebox[0mm][l]{$\hspace{-15\pmm}\scriptstyle u\ \  \ \,\,\,x\, y\, z$} 
}
=\int_{k,p} \ \bn\rme^{i k r(u)}\en\ \bn\rme^{-ikr(x)}\en
 \ \bn\rme^{i p r(y)}\en\  \bn\rme^{-ipr(z)}\en \nn\\
 &&= 
 \int_{k,p} \ \bn\rme^{i k r(u)}\en\ \bn\rme^{-ikr(x)}\,
 \rme^{i p r(y)}\,\rme^{-ipr(z)}\en
  \rme^{-p^2 C(y-z)} \, \rme^{kp[C(x-z)-C(x-y)]}
\ .\nn
\eea
We want to study the contraction of $x$, $y$, and $z$, and look for all 
contributions which are proportional to $
\GB= \int_{k} \,\bn\rme^{i k r(u)}\en\  \bn\rme^{-ikr((x+y+z)/3)}\en$.
The key-observation is that in \Eq{m:uiui} the leading term is obtained
by approximating $\rme^{kp[C(x-z)-C(x-y)]}\approx 1$. 
All subsequent terms yield factors of $k$, which  after integration 
over $k$ give  derivatives of the $\tilde \delta^d$-distribution.
The result is that 
\be\label{m:no div}
\MOPE \FD \GB - \MOPE{\GH} \1 =0 \ . 
\ee
This means that divergences of $\FD$ are already taken into account 
by a proper treatment of the divergences in $\GH$, analyzed in 
\Eq{m:symbmope1}.

\section{Renormalization at 1-loop order}
\label{m:Renormalization at 1-loop order}
In the last sections, we discussed how divergences occur, 
how their general structure is obtained by the MOPE, and how 
the MOPE-coefficients are calculated. 
In the next step, the theory shall be renormalized.
The basic idea is to identify the divergences through the MOPE,
and then to introduce
counter-terms which  subtract these divergences.
These counter-terms are nothing else than 
integrals over the MOPE-coefficients, properly regularized, i.e.\ %
cut off. 
We introcuce two 
renormalization group factors $Z$ (renormalizing the field $r$) 
and $Z_b$ (renormalizing the coupling $b$). Recalling 
\Eq{m:2ptact}, this is 
\bea \label{m:Ham1}
{\cal H}[ r]&=& \frac{Z}{2-D}\int_x \half 
\big(\nabla r(x)\big)^2 + b Z_b \mu^\varepsilon \int_x \int_y
\tilde \delta^d\big ( r(x)-  r(y)\big ) \ ,
\eea
where $r$ and $b$ are the renormalized field and renormalized 
dimensionless 
coupling constant,
and $\mu=L^{-1}$ is the renormalization momentum scale.
(To be precise: The field $r$ in \Eq{m:2ptact} is the bare field 
and should be noted $r_0$.)

Let us start to eliminate the divergences in the case,
where the  end-points $(x,y)$  of a single dipole 
are contracted towards  a point
(taken here to be the center-of-mass $z=(x+y)/2$).  The MOPE
was given in \Eq{m:symbmope1} ff. 
We there have to distinguish between counter-terms for 
relevant operators and those for marginal operators.
The former can be defined by analytic continuation, while the latter require
a subtraction scale. 
Indeed, the divergence proportional to $\mbox{\bf 1}$ is given by the integral
\be
	\int_{\Lambda^{-1}< |x-y| < L} 
	\MOPE{\,\GHxyd\,}{\1}
	=\int_{\Lambda^{-1}}^{L} \frac{\rmd x}{x} x^{D-\nu d}
	=\frac1{D-\E} \left( \Lambda^{D-\E} -L^{\E-D}\right)
\ ,
\ee
where $\Lambda$ is a high-momentum UV-regulator and $L$ a large distance
regulator.
For $\E\approx 0$ this is UV-divergent but IR-convergent.
The simplest way to subtract this divergence is therefore to replace the
dipole operator by 
\be
\strut_x\GB\strut_y\  \longrightarrow\  \strut_x\GB\strut_y\,-\,
\strut_x\GBdotted\strut_y \ ,
\ee
where $\strut_x\GBdotted\strut_y = |x-y|^{-\nu d} $. 
This amounts to adding to the bare Hamiltonian (\ref{m:2ptact}) the UV-divergent
counter-term
\be \label{m: CT for 1 in action}
\Delta{\cal H}_{\1}=-b Z_b \mu^\E \,	\int_x \int_y |x-y|^{-\nu d} \ , 
\ee
which is a pure number and thus does not change the expectation-value of any
physical observable.

We next consider marginal operators:
In the MOPE of \Eq{m:symbmope2},
the integral over the relative distance of 
${\int_{x-y}}\MOPE{\,\GHxyt\,}{\GOalphabeta}  \GOalphabeta $
is logarithmically divergent at $\E=0$.
In order to find the appropriate counter-term, we use dimensional
regularization, i.e.\ set $\E>0$.
An IR-cutoff $L$, or equivalently a subtraction momentum scale $\mu=L^{-1}$,
has to be introduced in order to define the subtraction operation.
{\em As a general rule, let us integrate over all distances appearing in the
MOPE-coefficient, bounded by the subtraction scale $L=\mu^{-1}$.} 
Defining
\vspace*{-2mm}
\be  \label{m: CT for + in action-}
\DIAGind \GH  \GOalphabeta{L} :=
\int_{|x-y|<L} \MOPE{\, \GHxyd \,}  \GOalphabeta \,
\ee
we need the following counter-term in the Hamiltonian 
\be \label{m: CT for + in action}
\Delta{\cal H}_{\GO} = 
	-b \mu^\E \,{ \DIAGind \GH  \GOalphabeta {L}}\,\int_x \GOalphabeta_x  \ ,
\ee
subtracting explicitly the divergence in the integrals. 
The reader is invited to verify this explicitly 
on 
the example of the expectation value of ${\cal O}=\rme^{ik[r(s)-r(t)]}$. 
The solution is given in  
appendix H of \cite{WieseHabil}.

Since the angular integration in \Eq{m: CT for + in action-}
reduces $\ds\GOalphabeta$ to $\ds\GO$,  
we can replace \Eq{m: CT for + in action} by the 
equivalent expression
\be
\Delta{\cal H}_{\GO} = 
	-b \mu^\E \,{ \DIAGind \GH  \GO {L}}\,\int_x \GO_x  \ ,
\ee
for which the numerical value of the diagram is calculated as
\be
\label{m:MOPEL}
\DIAGind \GH  \GO{L} =
\int_{|x-y|<L} \MOPE{\, \GHxyd \,}  \GO 
 = -\frac1{2D} \int_0^L \frac{\rmd x}x x^{2D-\nu d} =-\frac{1}{2D} \frac
 {L^\E}{\E} \ .
\ee 
We can now subtract this term in 
 a minimal subtraction scheme (MS).
The internal dimension of the membrane $D$ is kept fixed and 
(\ref{m:MOPEL}) is expanded as a Laurent series in $\varepsilon$, which here
starts at $\varepsilon^{-1}$.
The residue of the pole in  \Eq{m:MOPEL} is
\vspace*{-3mm}
\be \label{m:<-|+> eps}
\DIAGind \GH  \GO{\E} = -\frac{1}{2D} \ .
\ee
It is this pole that is  subtracted in the MS-scheme by adding to the 
Hamiltonian a counter-term 
\be \label{m:CT number}
\Delta{\cal H}_{\GO}=-\,\frac b \E \,\DIAGind \GH  \GO{\E}  \,
\int_x \GO_x \ .
\ee

Similarly, the divergence arising from the contraction of two dipoles to 
a single dipole is subtracted by a counter-term
\be
\label{m:CTb integral}
\Delta{\cal H}_{\GB}=b^2\mu^{2\E}\,\DIAGind \GM  \GB L\,
\int_x\int_y {}_x\GB_y \,\ ,
\ee
with 
\be
\DIAGind \GM  \GB L =
	\int_{|x|<L}\,\int_{|y|<L} \MOPE \GM  \GB 
\ .
\ee
Reducing this integral counter-term to a number, we subtract the 
residue of the single pole of
\be \!\,
\DIAGind \GM  \GB L =\!
	\int\limits_{|x|<L}\,\int\limits_{|y|<L} \MOPE \GM  \GB = \!\int\limits_{|x|<L}\,\int\limits_{|y|<L}\left( |x|^{2\nu} {+} |y|^{2\nu} \right)^{-d/2}.
\label{m:MOPE2L}
\ee
Note that the regulator $L$ cuts off both  integrations. 
One can now either utilize some simple algebra or show 
by the  methods of conformal mapping (see \cite{WieseHabil})
that the residue is obtained by fixing one distance to
1 and by freely integrating over the remaining one \vspace*{-3mm}
\be \label{m:<=|-> eps}
\DIAGind \GM  \GB {\E} =\int_0^\infty \frac{\rmd x}x x^{D} 
\BRA{1+x^{2-D}}{-\frac{2D}{2-D}} 
= \frac1{2-D} \frac{\Gamma\left(\frac D{2-D}\right)^{\!2}}
{\Gamma\left(\frac{2D}{2-D}\right)}
\ .
\ee
As a result,  the model is UV-finite at 1-loop order, if we use in the
renormalized Hamiltonian \eq{m:Ham1}
the renormalization factors $Z$ and $Z_b$
\bea 
Z&=& 1-(2-D)\DIAGind \GH  \GO{\E} \,\frac{b}\E \,+\,{ O}(b^2)
\label{m:Z1}
\\
Z_b&=&1+\DIAGind \GM  \GB {\E} \,\frac b\E \,+\,{ O}(b^2)
\label{m:Zb1}
\ .
\eea
Note that due to \Eq{m:no div} no counter-term for $\FD$ is necessary. 

The renormalized field and coupling are re-expressed in terms of their bare 
counterparts through 
\be \label{m:bare ren}
r_0(x)=Z^{1/2}\,r(x)
\ ,\qquad
b_0=b\,Z_b\,Z^{d/2}\,\mu^\E
\ .
\ee
Finally,
the renormalization group functions  are obtained from the variation of the
coupling constant and the field with respect to the renormalization scale $\mu$,
keeping the bare coupling fixed.
The flow of the coupling is written in terms of $Z$ and $Z_b$ 
with the diagrams given in \Eqs{m:<-|+> eps} and \eq{m:<=|-> eps}
as
\bea
\label{m:beta}
\beta(b) &:=&
\mu \AT{\frac{\partial}{\partial \mu }}{b_0} b
=\frac{-\E b} {1+ b\frac{\partial}{\partial b } \ln Z_b +
\frac{d}2 b \frac{\partial}{\partial b} \ln Z}\nn\\
&=& -\E b +\left( \DIAGind \GM  \GB{\E} - 
\nu d \DIAGind \GH  \GO{\E} \right) b^2
+{ O}(b^3) 
\ .
\eea
Similarly, the full dimension of the field (the exponent entering
into the correlation function) is obtained as
\bea
\label{m:nu}
\nu (b) &:=&
\frac{2-D}{2}-\half \AT{\mu \frac{\partial}{\partial \mu }}{b_0} \ln Z =
\frac{2-D}2 -\half \beta(b) \frac{\partial}{\partial b} \ln Z\nn\\
&=& \frac{2-D}2 \left[ 1 -b \DIAGind \GH  \GO {\E} \right]
+{O} (b^2) =\frac{2-D}2 \left[ 1 + b \frac1{2D} \right]
+{O} (b^2)  \ .\qquad
\eea
Note that minimal subtraction is used on the level of counter-terms
or equivalently $Z$-factors. Since $Z$ enters as $Z^d$ into the
$\beta$-function, the latter also contains a factor of $d$ in the
1-loop approximation, i.e.\ $Z^d$ is not minimally renormalized. 
In order to calculate the leading order in 
$\E$, the factor of  $d$ can be replaced by $d_c=\frac{4D}{2-D}$. 

The $\beta$-function has a non-trivial  fixed-point with
$\beta(b^*)=0$, which has positive slope and thus describes the 
behavior of the model at large distances.
The anomalous dimension $\nu^*:=\nu(b^*)$ becomes to first order in 
$\E$
\be
\nu^*=\frac{2-D}2 \bigg[ 1 +  \frac\E{2D} \left({\tx\frac1{2-D} \frac{\Gamma\BRA{\frac D{2-D}}{2}}
	{\Gamma\left(\frac{2D}{2-D}\right)}+1}\right)^{\!\!-1}\bigg] +{ O}(\E^2)\ .
\ee
For polymers, this result reduces to the well-known formula
$ \nu^*(D=1) = \half+ \frac{4-d}{16} +{ O}((4-d)^2)$.

\section{Results for self-avoiding membranes from 2-loop calculations}
Two loop results  are obtained by an explicit analytic and numeric integration
of the following combination of diagrams (which are all of 
order $1/\E$)
\bea 
\hspace{-1cm}
{\cal C}_1 &=& -\frac23 \DIAGind \GJ  \GB {L} + 
\left(\DIAGind \GM  \GB {L}\right)^2 \label{2:C1}\nn\\
\hspace{-1cm}
{\cal C}_2 &=&  -2 \DIAGind \GP  \GB {L} +
\DIAGind \GH  \GO {L} \DIAGind \GX  \GB {L} 
\label{2:C2}\nn\\
\hspace{-1cm}
{\cal C}_3 &=& \DIAGind \GH  \GO {\E^{-1}} \left(  \DIAGind \GX  \GB {L} 
+{(2-D) d \over 4} \DIAGind \GM  \GB {\E^{-1}} \right)
\label{2:C3}\nn\\
{\cal F}_1&=&\half \DIAGind \GI  \GO {L} {-}\half 
\DIAGind \GH  \GO {L}	 
\DIAGind \GZ  \GO {L} {-} \half \DIAGind \GM  \GB {L} 
\DIAGind \GH  \GO {L} 
\label{2:F1}\nn\\
{\cal F}_2&=&  -\half \DIAGind \GH  \GO {\E^{-1}} \times
\weiter \times \left(  \DIAGind \GZ  \GO {L} +
\frac{(2-D)(d+2)}2  \DIAGind \GH  \GO {L}
-\DIAGind \GH  \GO {L} \DIAGind \GW  \GO {L} \right)  \nn\\
{\cal F}_3&=&\half \DIAGind \GH  \GO {\E^{-1}} 
\left( \DIAGind \GM  \GB {L}
-\DIAGind \GM  \GB {\E^{-1}}
\right)
\label{2:F3}
\ .
\eea 
\begin{figure}[t] 
\centerline{
\epsfxsize=0.68\textwidth \parbox{0.68\textwidth}
{\epsfbox{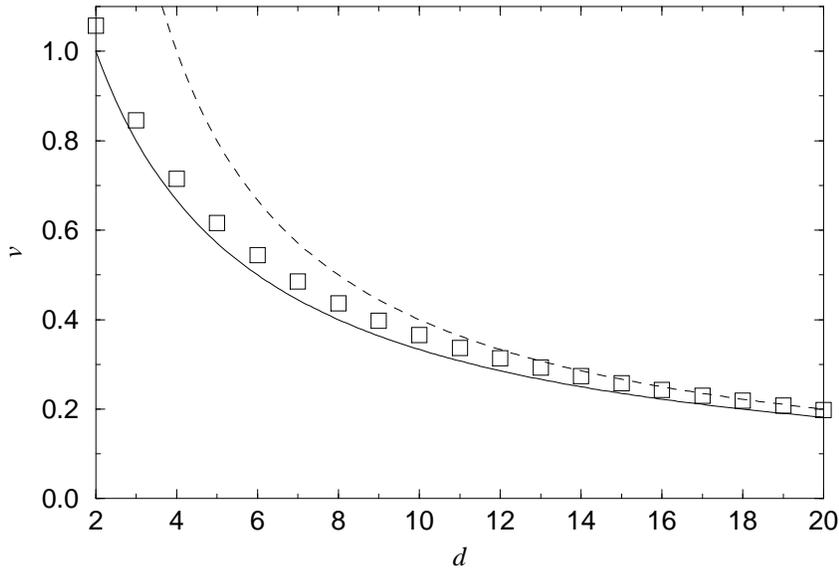}}}
\caption{Extrapolation of the 2-loop results in $d$ and $\E$ for membranes
$D=2$ in $d$ dimensions, using the expansion of $\nu^*(d+2)$ (squares). The solid line is the prediction made
by Flory's approximation, the dashed line by the variational ansatz.}
\label{x:f:nu(d)}
\end{figure}%
One then has the freedom to extrapolate about any point on 
the critical curve $\E(D,d)=0$. This freedom -- or ambiguity -- can 
be used to optimize the results. 
The results of such a procedure for the
 2-loop cacluations of $\nu^*$ are given
 on \Fig{x:f:nu(d)}
for membranes ($D=2$) in $d$ dimensions ($2\le d\le 20)$.
We see that for $d\to\infty$ the prediction of a Gaussian variational
method $
\nu_{\ind{var}}=\frac{2D}{d}$
becomes exact.
For small $d$, the prediction made by Flory's argument 
$\nu_{\ind{Flory}}=\frac{2+D}{2+d}$
is close to our
results.
This is a non-trivial result, since the membrane case corresponds to
$\E=4$ and in comparison with polymers in $d=3$, where $\E=1/2$,  the 2-loop corrections
were expected to be large.
In fact,  they are small when one expands around
the critical curve $\E=0$ for an adequate range of $D\sim 1.5$ (depending
slightly on $d$ and on the choice of variables) and a suitable
choice of extrapolation variables.
In this case the 2-loop corrections are even smaller than the 1-loop
corrections and allow for more reliable extrapolations to $\E=4$.
This can be understood from the large order behavior
\cite{DavidWiese1998}.

Let us now turn to the physically relevant case of membranes in three
dimensions  ($D=2$, $d=3$). Our calculations predict an exponent
$\nu^*\approx 0.85$ or equivalently a fractal dimension of 
$d_\ind f \approx 2.4$, 
which is in agreement with those experiments and simulations
which find a fractal phase. However, this is still under debate,
and a lot of evidence has been collected that the flat phase
is generic. (For a more detailed discussion of this problem, 
see the review \cite{WieseHabil}.)

\section{Outlook}
In these lecture notes, we have only been able to discuss 
the simplest applications of non-local field theories.
Much more has been achieved during the last years. 

First of all, it has been demonstrated that the dynamics, described by a Langevin-equation
\be
\frac{\rmd}{\rmd t} r(x,t) = - \frac{\p{\cal H}[r]}{\p r(x,t)}+
\zeta(x,t) \ , \qquad \left< \zeta(x,t) \zeta(x',t')\right> = \lambda
\delta^D(x-x') \delta(t-t')
\ee
also leads to a renormalizable field theory \cite{Wiese1997a},
where the dynamic exponent $z$, defined by the decay of the auto-correlation
function $\left< [r(x,t)-r(x,t')]^2\right> \sim |t-t'|^{2/z}$ is given 
by $z=2+D/\nu^*=2+d_\ind f$, where $d_f=D/\nu^*$ is the fractal dimension
of the membrane. This result had been suggested long-time ago 
by DeGennes \cite{DeGennes1976a}
for polymers and by Kardar et al.\ for 
membranes \cite{KantorKardarNelson1986b},
but only with  the methods discussed above, a proof of that conjectures
could be given \cite{Wiese1997a}. 
When hydrodynamics is included, the dynamical exponent $z$ changes
to $z=d$ \cite{Wiese1997b}.

Interestingly, also disorder can be treated via the same methods,
since averaging over disorder leads to interactions very similar 
to self-avoidance. The above methods have been applied to 
the motion of a polymer or membrane in a static 
disordered force field with both potential and non-conserved
parts -- leading to new universal physics
\cite{LeDoussalWiese1997,WieseLedoussal1998}. 

Also the advection of a polymer in a turbulent flow has been
analyzed \cite{Wiese1999}, paralling the discusion of the passive advection of
particles, which in the turbulence community is known as the 
passive scalar problem. 

Another interesting generalization is to anisotropically tethered
membranes \cite{BowickFalcioniThorleifsson1997,%
BowickGuitter1997,BowickTravesset1998b}. In these
models, the membrane is more rigid in one direction,
forming a tubular phase.

One of the most useful tricks for self-avoiding polymers is
the mapping onto a massive scalar field-theory,  i.e.\ a $\phi^4$-theory
\cite{DeGennes1972}.
In the limit, where the number of compoents $n\to0$, results for 
self-avoiding polymers are obtained. Stated differently, 
$\phi^4$-theory is a generalization of self-avoiding polymers. 
Another generalization has been discussed above: The generalization 
to membranes with internal dimension $D\not= 1$. The question arizes, 
whether a common generalization of both the $\phi^4$-theory and 
polymerized membranes is possible. Such a model has indeed
been constructed \cite{WieseKardar1998a,WieseKardar1998b}, and 
leads to interesting new physics. 

Another still open question is the analysis of the spectrum of 
subdominant operators \cite{WieseShpot1998}. This may give a hint
of why polymerized membranes are generically seen flat in simulations.

\section{Acknowledgements}
It is a pleasure to thank the organizers of the 
workshop at Heidelberg, Hans J\"urgen Pirner and 
Franz Wegner, for the opportunity to give these lectures.
I wish the honoree of this volume 60 more years of fruitful scientific
work.

\nocite{DDG1,DDG2,Duplantier1987,DuplantierHwaKardar1990}


\end{document}